\documentclass[9pt,conference]{IEEEtran}
\usepackage{waspaa25} % removes hyperlinks (required by IEEE Xplore)

\usepackage{cite}
\usepackage{amsmath,amssymb,amsfonts}
\usepackage{algorithmic}
\usepackage{graphicx}
\usepackage{textcomp}
\usepackage{xcolor}
\usepackage{epsf}
\usepackage{times}
\usepackage{float}
\usepackage{wasysym}
\usepackage{color}
\usepackage{stfloats}
\usepackage{subeqnarray}
\usepackage{multirow}
\usepackage{booktabs}
\usepackage{subcaption}
\usepackage{url}
\usepackage{tablefootnote}
\usepackage{hyperref}

\title{Multi-Utterance Speech Separation and Association \\ Trained on Short Segments}

\name{Yuzhu Wang$^{1}$,
      Archontis Politis$^{1}$,
      Konstantinos Drossos$^{2}$,
      Tuomas Virtanen$^{1}$}
\address{$^{1}$Signal Processing Research Center, Tampere University, Tampere, Finland \;
$^{2}$Nokia Technologies, Espoo, Finland
}

\begin{document}

\maketitle

\begin{abstract}
Current deep neural network (DNN) based speech separation faces a fundamental challenge --- while the models need to be trained on short segments due to computational constraints, real-world applications typically require processing significantly longer recordings with multiple utterances per speaker than seen during training. 
In this paper, we investigate how existing approaches perform in this challenging scenario and propose a frequency-temporal recurrent neural network (FTRNN) that effectively bridges this gap.
Our FTRNN employs a full-band module to model frequency dependencies within each time frame and a sub-band module that models temporal patterns in each frequency band. 
Despite being trained on short fixed-length segments of 10~s, our model demonstrates robust separation when processing signals significantly longer than training segments (21-121~s) and preserves speaker association across utterance gaps exceeding those seen during training.
Unlike the conventional segment-separation-stitch paradigm, our lightweight approach (0.9~M parameters) performs inference on long audio without segmentation, eliminating segment boundary distortions while simplifying deployment.
Experimental results demonstrate the generalization ability of FTRNN for multi-utterance speech separation and speaker association.
\end{abstract}


\vspace{1ex}
\section{Introduction}
\vspace{1ex}
Speech separation aims to isolate individual speech signals from mixtures containing multiple speakers.
Current deep neural network (DNN) based separation models need to be trained on short segments~\cite{vincent2018audio, purwins2019deep, cobos2022overview, wang2018supervised, luo2019conv, Yu2017permutation, kolbaek2017multitalker, Wang2023TFGridNet, Wang2023TFGridNet2, Hershey2016, wang2019low}.
However, real-world applications reveal significant performance degradation when these models process longer recordings~\cite{Zeghidour2020, chen2020continuous, li2021dualmodeling}.
This generalization issue remains underexplored in existing literature. Even the methods that perform excellently on short recordings struggle when applied to long speech separation --- separating mixture signals consisting of multiple utterances per speaker into individual streams, ensuring all utterances from the same speaker are consistently associated to the same output stream.

\begin{figure}
\centering
\includegraphics[width=0.48\textwidth]{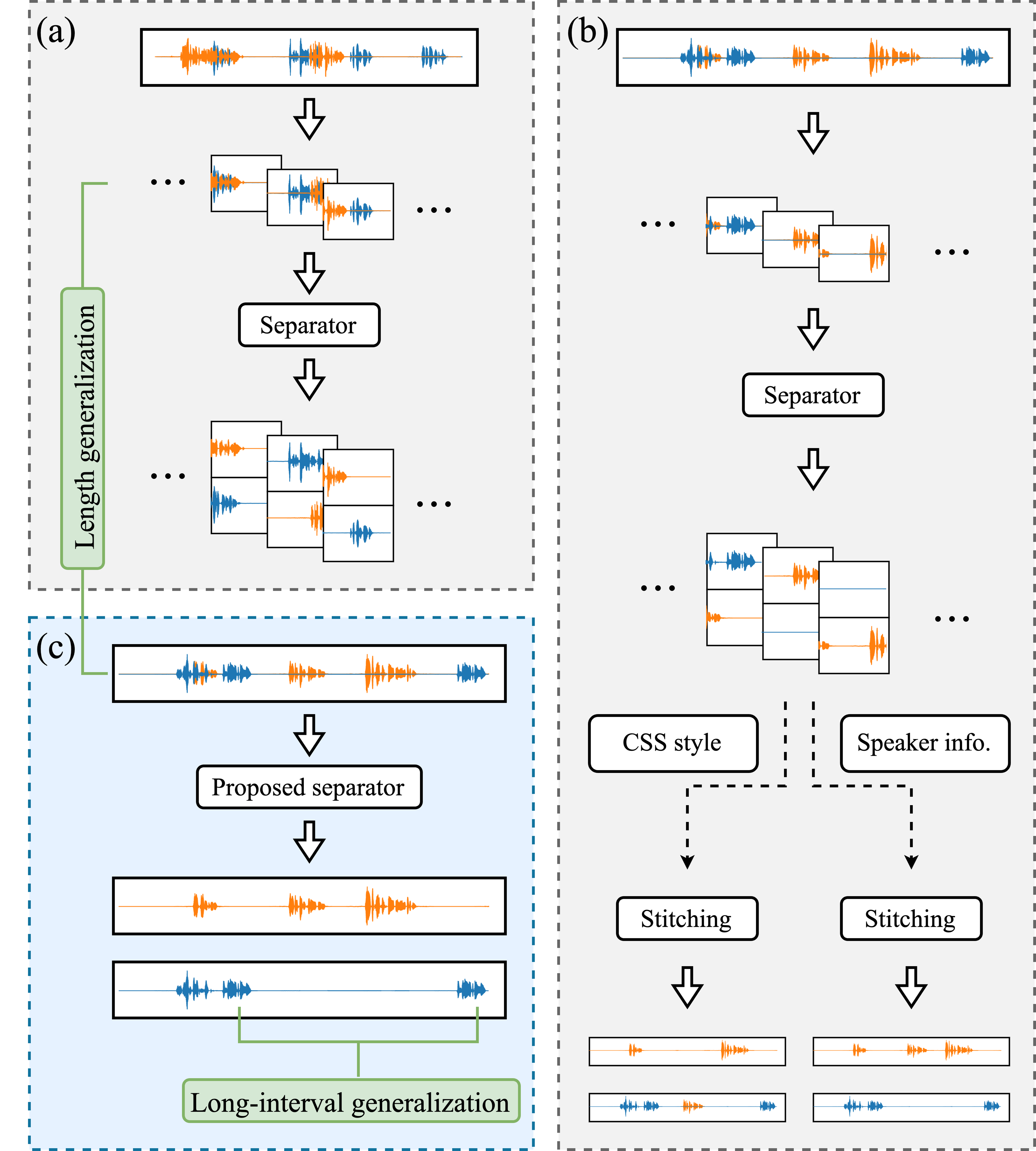}
% \vspace{-.5em}
\caption{Visualization of long speech separation: (a) training phase on short segments, (b) segment–separate–stitch inference, (c) our proposed recurrent neural network inference on unsegmented mixture.}
\label{fig-problem}
% \vspace{-.6cm}
\end{figure}

Current approaches typically employ a segment-separation-stitch paradigm to address long speech separation, as shown in Fig.~\ref{fig-problem}(b), where long recordings are split into short segments, processed independently, and then stitched.
This processing paradigm does not leverage contextual information across segments.
When implementing this paradigm, the challenge lies in determining the correct permutation of separated segments during stitching, particularly when utterances from the same speaker are separated by long intervals.

Continuous speech separation (CSS)~\cite{chen2020continuous, li2021dualmodeling, han2021continuous, li2021dual, luo2020dual, von2022segment} follows this paradigm by stitching separated segments into output streams, with each stream potentially containing utterances from different speakers rather than isolating individual speakers, as shown in Fig.~\ref{fig-problem}(b).
CSS primarily aims to reduce the overlap rate in output streams.
% rather than addressing permutation ambiguity or speaker association. 
To achieve CSS-style outputs, the approaches require very short segments, ensuring that each segment contains no more than $2$-$3$ speakers~\cite{li2021dualmodeling}. 
This constraint leads to frequent stitching operations, increasing the risk of introducing distortions at segment boundaries.

An alternative solution leverages speaker-wise representation in separating long recordings, as shown in Fig.~\ref{fig-problem}(b).
These methods incorporate speaker modeling techniques such as speaker clustering~\cite{Zeghidour2020, Hershey2016, wang2019low}, speaker identification~{\mbox{\cite{kenny2013plda, snyder2018x, snyder2019speaker}}}, or attractor-based mechanisms~\cite{Chen2017deep, Luo2018speaker, Chetupalli2022, lee2024boosting, wang2025attractor} to determine the optimal assignment of separated segments.
While effective, these methods often require speaker modeling components or auxiliary speaker identification modules, increasing implementation complexity.

In this paper, we investigate how speech separation models trained on short fixed-length segments perform when applied to long recordings with multiple utterances per speaker. Our main contributions are twofold. 
First, we evaluate the generalization performance of existing speech separation methods when processing recordings significantly longer than their training segments, an aspect previously underexplored in literature despite its importance for real-world applications.
Our analysis reveals that current methods show limited performance in long speech separation, particularly when utterances from the same speaker are separated by extended silent intervals.
Second, we propose a frequency-temporal recurrent neural network (FTRNN) that effectively bridges the gap between fixed-length training and flexible-length inference.
As illustrated in Fig.~\ref{fig-problem}(a) and Fig.~\ref{fig-problem}(c), our proposed FTRNN offers two key advantages: 
(1) though trained on fixed-length segments, it generalizes to flexible-length inputs during inference without performance degradation; (2) while trained with short utterance intervals, it maintains speaker association across substantially longer intervals than seen during training.
Experimental results on our generated multi-utterance dataset demonstrate that the proposed lightweight FTRNN ($0.9$~M parameters) significantly outperforms existing methods across various utterance gaps and numbers.

\vspace{1ex}
\section{Signal Model}
\label{sec-signal-model}
\vspace{1ex}
A single-channel mixture signal can be formulated in the time domain as
% \vspace{-1ex}
\begin{equation}
\label{eq:observed_signal}
y(t) = \sum_{c=1}^{C} x_c(t) + n(t).
% \vspace{-1mm}
\end{equation}
Here, the observed signal $y(t)$ consists of speech signals $x_c(t)$ from $C$ speakers and noise $n(t)$. The $x_c(t)$ represents the time-domain signal from the $c$-th speaker, which consists of multiple utterances and potential pauses between them. 
Our objective is to perform speaker-independent separation to estimate $x_c(t)$, separating all utterances belonging to the same speaker into a single output, where the output indices of speakers can be arbitrary, as illustrated in Fig.~\ref{fig-problem}(c). In this study, we focus on the two-speaker case where $C=2$.

\vspace{1ex}
\section{Proposed method}
\label{sec-proposed-system}
\vspace{1ex}
The proposed FTRNN is trained on fixed-length short segments but generalizes to flexible-length long audio during inference without requiring any segmentation operations, as illustrated in Fig.~\ref{fig-problem}(a) and Fig.~\ref{fig-problem}(c).
During training, we split long audio into $10$-second segments as input signals. 
The FTRNN processes these segments independently, separating each into two output channels, with each channel containing all utterances from a single speaker. 
We employ permutation invariant training (PIT)~\cite{kolbaek2017multitalker} with the scale-invariant signal-to-distortion ratio (SI-SDR)~\cite{LeRoux2018a} loss function, where the PIT is applied to the separated segment pairs independently.
Notably, there is no need to enforce consistent speaker ordering across segments during training, meaning the same speaker might appear in different output channels for different segments.

\begin{figure}
\centering
\includegraphics[width=0.45\textwidth]{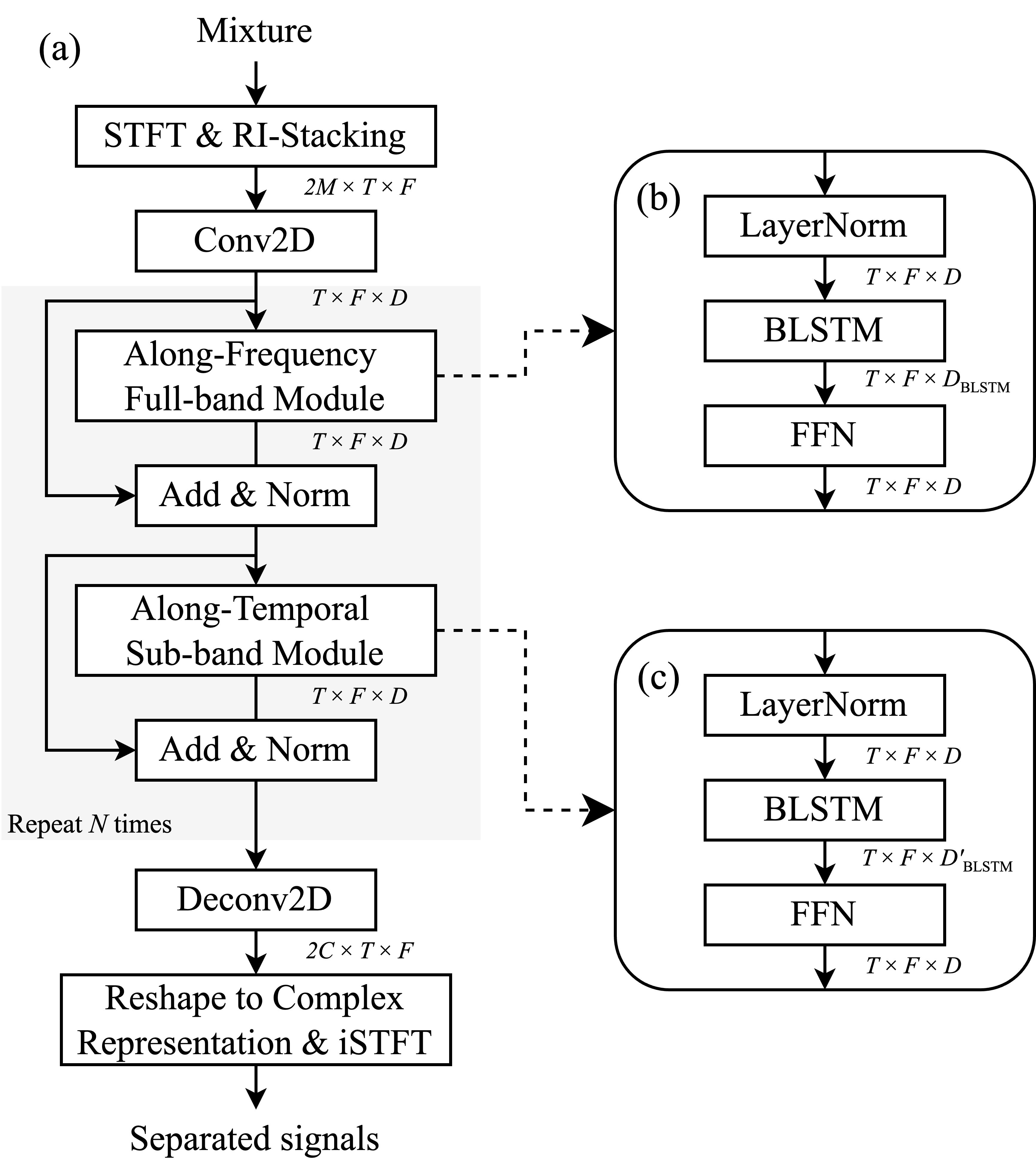}
% \vspace{-.5em}
\caption{Proposed system architecture.}
\label{fig-system}
% \vspace{-.5cm}
\end{figure}

The architecture of FTRNN is shown in Fig.~\ref{fig-system}.
The input mixture is transformed into the time-frequency domain via short-time Fourier transform (STFT), with real and imaginary components stacked to form a tensor of shape $2 \times T \times F$. Here, $T$ is time frames and $F$ is frequency bins.
A 2D convolutional layer with a $3 \times 3$ kernel transforms the stacked tensor into a latent representation of shape $T \times F \times D$, where $D$ is the feature dimension.
This representation then passes through our full-band and sub-band modules.

\textit{Along-Frequency Full-band Module:} The along-frequency full-band module is shown in Fig.~\ref{fig-system}(b). After layer normalization, a bidirectional long short-term memory (BLSTM) network operates along the frequency dimension for each time frame independently. The BLSTM captures both forward and backward dependencies across frequencies, where $D_{\text{BLSTM}}$ is the hidden dimension. The BLSTM output is then projected back to the feature dimension $D$ through a feed-forward network (FFN).

\textit{Along-Temporal Sub-band Module:} The along-temporal sub-band module in Fig.~\ref{fig-system}(c) focuses on temporal modeling. 
After layer normalization, a BLSTM network operates along the temporal dimension for each frequency bin independently, where $D'_{\text{BLSTM}}$ is the hidden dimension.
A FFN then maps the BLSTM outputs back to the feature dimension $D$. 

Both modules employ BLSTM networks, which are key to the model's ability to process signals longer than the segments used during training.
The two modules are arranged sequentially with residual connections and layer normalization, repeated $N$ times. 
Finally, a 2D deconvolutional layer maps the processed features back to the time-frequency domain with $2C$ output channels, where $C$ is the number of speakers. After reshaping to the complex representation, an inverse STFT (iSTFT) transforms the output back to the time domain, yielding $C$ separated signals.

\vspace{1ex}
\section{Experimental settings}
\label{sec-sxperimental-settings}
\vspace{1ex}
% \vspace{-1ex}
\vspace{1ex}
\subsection{Datasets}
\label{ssec-datasets}
\vspace{1ex}
For the multi-utterance scenario, we generated a dataset using speech from the LibriSpeech corpus~\cite{panayotov2015librispeech} and noise from the DEMAND dataset~\cite{thiemann2013diverse}.
For each mixture, we randomly selected two speakers from LibriSpeech. For each selected speaker, we randomly extracted between 4 and 5 speech utterances. We preserved the original Librispeech signal levels without scaling, with relative levels between any two speech utterances varying within $[0, 5]$ dB. 
We inserted random silence intervals at the beginning and between each utterance, with durations uniformly sampled from $[1, 3]$ seconds. Concatenating all speech utterances and silences for each speaker yielded a complete speaker signal. 

To simulate reverberation, we randomly generated room parameters: dimensions ranging from $[4.0, 8.0]$~m in length and width, and $[3.0, 4.0]$~m in height. The RT${60}$ values varied between $0.2$ and $0.6$ seconds. A single microphone was randomly positioned at heights from $1.0$ to $1.5$~m, while speakers were randomly generated at heights between $1.5$ and $2.0$~m. Microphone and speakers were restricted to be at least $0.5$~m away from any walls and each other.  
Using the gpuRIR toolkit~\cite{diaz2021gpurir}, we applied reverberation to each speaker individually.
\begin{table}[tpb]
\setlength\tabcolsep{3pt}
\renewcommand{\arraystretch}{1.0}
\caption{Test datasets with varying utterance configurations.}
% \vspace{-1.5ex}
\label{table_testset}
\centering
\resizebox{0.99\linewidth}{!}{
\begin{tabular}{l|cccccc}
\toprule
\multirow{2}{*}{\#} & \multirow{2}{*}{Samples} & Utterances & Utterance & Average & Minimum & Maximum  \\
 & & per speaker & gap (s) & length (s)  & length (s) & length (s) \\
\midrule
0 & 2000 & [4, 5] & [1, 3]  & 55.2 & 21.3 & 121.2 \\
\midrule
1 & 2000 & 2 & [2, 4]  & 24.5 & 10.3 & 60.5 \\
2 & 2000 & 2 & [9, 11]  & 33.2 & 18.3 & 65.8 \\
3 & 2000 & 2 & [19, 21]  & 43.3 & 27.8 & 77.4 \\
4 & 2000 & 2 & [29, 31]  & 53.2 & 38.3 & 86.7 \\
5 & 2000 & 2 & [39, 41]  & 63.1 & 46.9 & 97.1 \\
\midrule
6 & 2000 & 1 & [2, 4]  & 11.3 & 3.4 & 29.2 \\
7 & 2000 & 3 & [2, 4]  & 36.2 & 15.3 & 78.9 \\
8 & 2000 & 4 & [2, 4]  & 48.3 & 22.1 & 98.6 \\
9 & 2000 & 5 & [2, 4]  & 60.7 & 28.5 & 119.4 \\
\bottomrule
\end{tabular}
}
% \vspace{-1ex}
\end{table}
To add noise, we randomly extracted segments of equal length from the DEMAND noise dataset and adjusted their amplitude according to a randomly generated signal-to-noise ratio (SNR) between $0$ and $10$~dB. In the SNR calculation, the signal level was computed as the average power of all speakers on a logarithmic scale.

All generated signals were single-channel with a sampling frequency of $16$~kHz. The training and test sets did not share speakers or noise signals to ensure proper evaluation. 
% The generated dataset has an average overlap ratio of $XXXXXX$~\%.
The generated dataset comprised $20000$ training samples, $2000$ validation samples, and $2000$ test samples.

The generated test set is shown in Table~\ref{table_testset} (test set \#~$0$). To evaluate generalization capabilities, we generated nine additional test subsets (\#~$1$-$9$) following the same steps but with varying utterance configurations, as shown in Table~\ref{table_testset}.

\vspace{1ex}
\subsection{Training Configurations and Evaluation Metrics}
\label{ssec-training-configurations}
\vspace{1ex}
The model was implemented using the ESPNet toolkit~\cite{watanabe2018espnet}. We employed the Adam optimizer with an initial learning rate of $1.0 \times 10^{-3}$. The training process was configured to run for a maximum of 200 epochs, with an early stop strategy. 
A batch size of $24$ was used.
All hyperparameters were determined through pre-experiments on the validation dataset.

For the STFT at a sampling rate of $16$~kHz, we used a Hann window with a window length of $512$ samples ($32$~ms) and a hop length of $256$ samples ($16$~ms). 
The feature dimension $D$ was established at $32$, and the number of repeated blocks was configured as $N=4$. 
The BLSTM layers in both the full-band module and the sub-band module were implemented with $96$ hidden units in each direction.
Gradient clipping was applied with an $L_2$ norm threshold of $5.0$. We did not incorporate dynamic mixing or data augmentation techniques. 

We use SI-SDR~\cite{LeRoux2018a} to evaluate separation quality and diarization error rate (DER)~\cite{bredin2017pyannote} to assess speaker activity and association.

% \vspace{-1ex}
\vspace{1ex}
\section{Results and Discussions}
\label{sec-Experimental-Results}
\vspace{1ex}

We employed DPRNN~\cite{luo2020dual}, DPTNet~\cite{Chen2020DPTnet}, SepFormer~\cite{Subakan2021} and TFGrid~\cite{Wang2023TFGridNet} as our baseline methods.
All baseline models were retrained on our generated multi-utterance dataset with their original implementations to ensure a fair comparison. 
For training, all models except TFGrid used a $10$~s segment size (TFGrid used $5$~s due to GPU memory constraints). For inference, DPRNN, DPTNet, and FTRNN can be evaluated both direct inference without segmentation and stitch inference, while SepFormer and TFGrid can only be evaluated with stitching inference due to high GPU memory requirements.

We evaluated the proposed FTRNN and baseline models using both direct inference and the segment-stitch paradigm with oracle stitching (representing the performance upper bound for this paradigm).
The stitching inference refers to the implementation in ESPnet \cite{watanabe2018espnet}, but with modifications to the permutation process. 
First, long audio recordings are divided into $5$~s segments with $20$~\% overlap between adjacent segments. Each segment is then processed independently through the trained model (FTRNN or baseline model).
Subsequently, we use ground truth reference signals to determine the optimal SI-SDR-based permutation for each segment pair.
Finally, the segments are stitched using overlap-and-add (averaging over the overlapped regions) to form complete separated signals.

\begin{table}[tpb]
\setlength\tabcolsep{3pt}
\renewcommand{\arraystretch}{1.0}
\caption{Comparison of overall performance (test set \#~$0$).}
% \vspace{-1.5ex}
\label{table_overall}
\centering
\resizebox{0.99\linewidth}{!}{
\begin{tabular}{lcccc}
\toprule
\multirow{2}{*}{Models} & \#Param & FLOPs & SI-SDR$\uparrow$ & DER\tablefootnote{\url{http://pyannote.github.io/pyannote-metrics}}$\downarrow$  \\
 & (M) & (G/s) & (dB) & (\%)  \\
\midrule
Unproc. & -- & -- & -1.2  & -- \\
\midrule
DPRNN\tablefootnote{\href{http://github.com/espnet/espnet/blob/master/espnet2/enh/separator/dprnn_separator.py}{espnet/espnet/blob/master/espnet2/enh/separator/dprnn\_separator.py}}~\cite{luo2020dual} & 2.6 & 21.1 & 11.6 & 9.1 \\
DPRNN (oracle stitching) & -- & -- & 11.4 & 10.3 \\
DPTNet\tablefootnote{\href{http://github.com/espnet/espnet/blob/master/espnet2/enh/separator/dptnet_separator.py}{espnet/espnet/blob/master/espnet2/enh/separator/dptnet\_separator.py}}~\cite{Chen2020DPTnet} & 2.6 & 21.7 & 8.3 & 17.4 \\
DPTNet (oracle stitching) & -- & -- & 8.4 & 15.8 \\
SepFormer\tablefootnote{\href{https://github.com/speechbrain/speechbrain/blob/develop/speechbrain/lobes/models/dual_path.py}{speechbrain/speechbrain/blob/develop/speechbrain/lobes/models/dual\_path.py}}~\cite{Subakan2021}~(oracle stitching) & 26.0 & 141.6 & 7.5 & 17.9 \\
TFGrid\tablefootnote{\href{https://github.com/espnet/espnet/blob/master/espnet2/enh/separator/tfgridnet_separator.py}{espnet/espnet/blob/master/espnet2/enh/separator/tfgridnet\_separator.py}}~\cite{Wang2023TFGridNet}~(oracle stitching) & 14.4 & 500.6 & 14.2 & 7.7 \\
\midrule
FTRNN & 0.9 & 27.8 & \textbf{15.2} & \textbf{6.9} \\
FTRNN (oracle stitching) & -- & -- & 14.6  & 7.0 \\
FTRNN (SA-SDR loss~\cite{von2022sa}) & -- & -- & 12.2  & 8.7 \\
\bottomrule
\end{tabular}
}
\vspace{1ex}
\begin{flushleft}
\footnotesize
\textit{Notes:} FLOPs are computed on 4-second segments and then divided by $4$ to yield values in Giga per second (G/s), where a single channel configuration with a batch size of $1$ and a sampling rate of $16$~kHz is used. We use the \textit{torch.utils.flop\_counter} for the FLOPs calculation.
% \vspace{-2ex}
\end{flushleft}
% \vspace{-5ex}
\end{table}
% \vspace{-1ex}
\vspace{1ex}
\subsection{Overall Performance}
\label{ssec-exp-overall}
\vspace{1ex}
Table~\ref{table_overall} compares our FTRNN with baseline systems on the generated multi-utterance dataset (test set \# $0$ in Table~\ref{table_testset}). 
% Training used $10$~s chunk size while inference was performed without chunking. 
The unprocessed mixture shows an SI-SDR of $-1.2$~dB. 
Among the baselines, TFGrid achieves the highest SI-SDR of $14.2$~dB with oracle stitching, while DPRNN offers the lowest parameter count at $2.6$~M and the most efficient computation ($21.1$~G/s). 

Our proposed FTRNN achieves $15.2$~dB SI-SDR and $6.9$~\% DER for direct inference, outperforming DPRNN (direct inference) by $3.6$~dB in SI-SDR and reducing DER by $2.2$~\%. 
When using stitching inference, FTRNN achieves $14.6$~dB SI-SDR, surpassing TFGrid (oracle stitching) by $0.4$~dB. 
The FTRNN contains $0.9$~M parameters, fewer than all baseline models.
When trained with SA-SDR~\cite{von2022sa} instead of SI-SDR, performance decreases to $12.2$~dB SI-SDR and $8.7$~\% DER. 

Compared to direct inference, DPRNN and FTRNN both show performance degradation with oracle stitching (DPRNN: $0.2$~dB SI-SDR drop; FTRNN: $0.6$~dB SI-SDR drop), while DPTNet exhibits a slight improvement ($0.1$~dB SI-SDR increase). 
These SI-SDR variations occur despite zero-error permutation alignment, likely because segmentation disrupts utterance continuity, introducing distortions at segment boundaries where utterances are split and later reconnected.
These results demonstrate that processing longer recordings directly is more effective than the segment-stitch approach, as longer segments allow the model to capture extended temporal context.

% \vspace{-1mm}
\vspace{1ex}
\subsection{Performance vs. Utterance Gap}
\label{ssec-exp-utt-gap}
\begin{figure}
\centering
\includegraphics[width=0.46\textwidth]{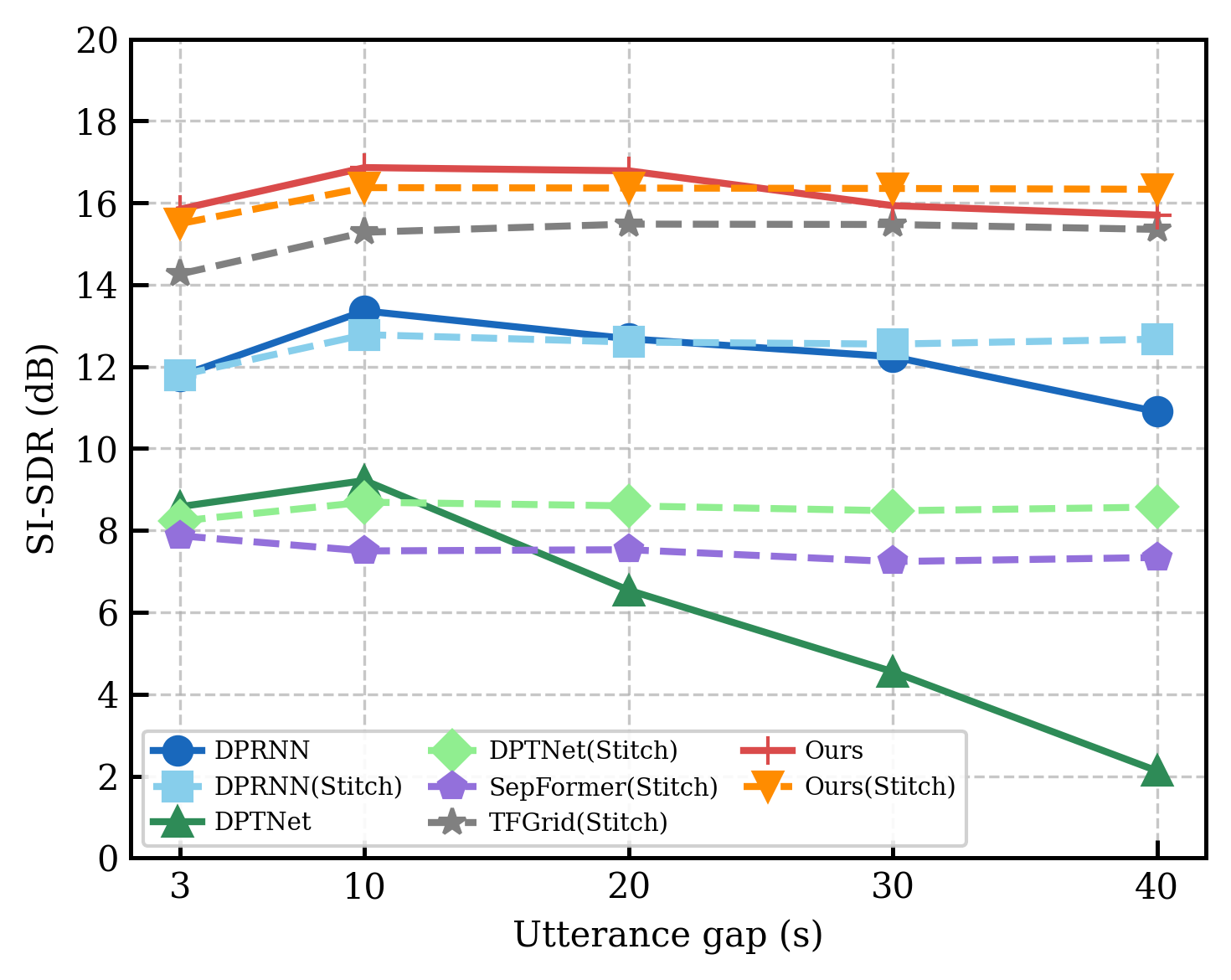}
% \vspace{-.5em}
\caption{SI-SDR performance comparison across different utterance gaps (test sets \# $1$-$5$ in Table~\ref{table_testset}). Each mixture contains two speakers with two utterances per speaker.}
\label{fig-utt-gap}
% \vspace{-.5cm}
% \vspace{-.5cm}
\vspace{1ex}
\end{figure}

\vspace{1ex}
We tested models on five test subsets to evaluate generalization performance across different utterance gaps (test sets \# $1$-$5$ in Table~\ref{table_testset}), with average gaps ranging from $3$~s to $40$~s.
Fig.~\ref{fig-utt-gap} shows the SI-SDR results across these test subsets. Notably, all models were trained using the training set with utterance gaps in the $[1, 3]$~s range, as described in Section~\ref{ssec-datasets}.

Our model maintains superior performance for direct inference across all gap durations, achieving $15.8$~dB SI-SDR at $3$~s gap and $15.7$~dB at $40$~s gap. This represents only a $0.1$~dB decrease despite increasing the gap by more than $13$ times.
In comparison, DPRNN drops from $11.8$~dB to $10.9$~dB ($0.9$~dB decrease), while DPTNet exhibits the largest degradation from $8.6$~dB to $2.1$~dB ($6.5$~dB decrease).
This performance degradation trend across models highlights the challenge of speaker separation and association when facing extended utterance gaps beyond training conditions.

With oracle stitching, DPRNN maintains $11.8$-$12.7$~dB, DPTNet stabilizes at $8.2$-$8.6$~dB, and TFGrid achieves $14.3$-$15.5$~dB. SepFormer shows minimal variation at $7.3$-$7.9$~dB. Our model with oracle stitching achieves $15.5$-$16.4$~dB.
These results reveal that even with limited generalization capability, models can achieve stable separation performance across long utterance gaps through stitching inference if an accurate permutation can be estimated.

% \vspace{-1ex}
% \vspace{-1mm}
\vspace{1ex}
\subsection{Performance vs. Utterances per Speaker}
\label{ssec-exp-utt-num}
\vspace{1ex}
To evaluate model generalization to unseen utterance counts, we used test sets \# $1$ and \# $6$-$9$ in Table~\ref{table_testset}, where the utterance gaps were uniformly sampled from $[2, 4]$~s range (average $3$~s).
During training, TFGrid used a $5$~s segment size, while all other models used a $10$~s segment size. 
Each segment usually contains parts of one or two utterances rather than complete ones.

\begin{figure}
\centering
\includegraphics[width=0.46\textwidth]{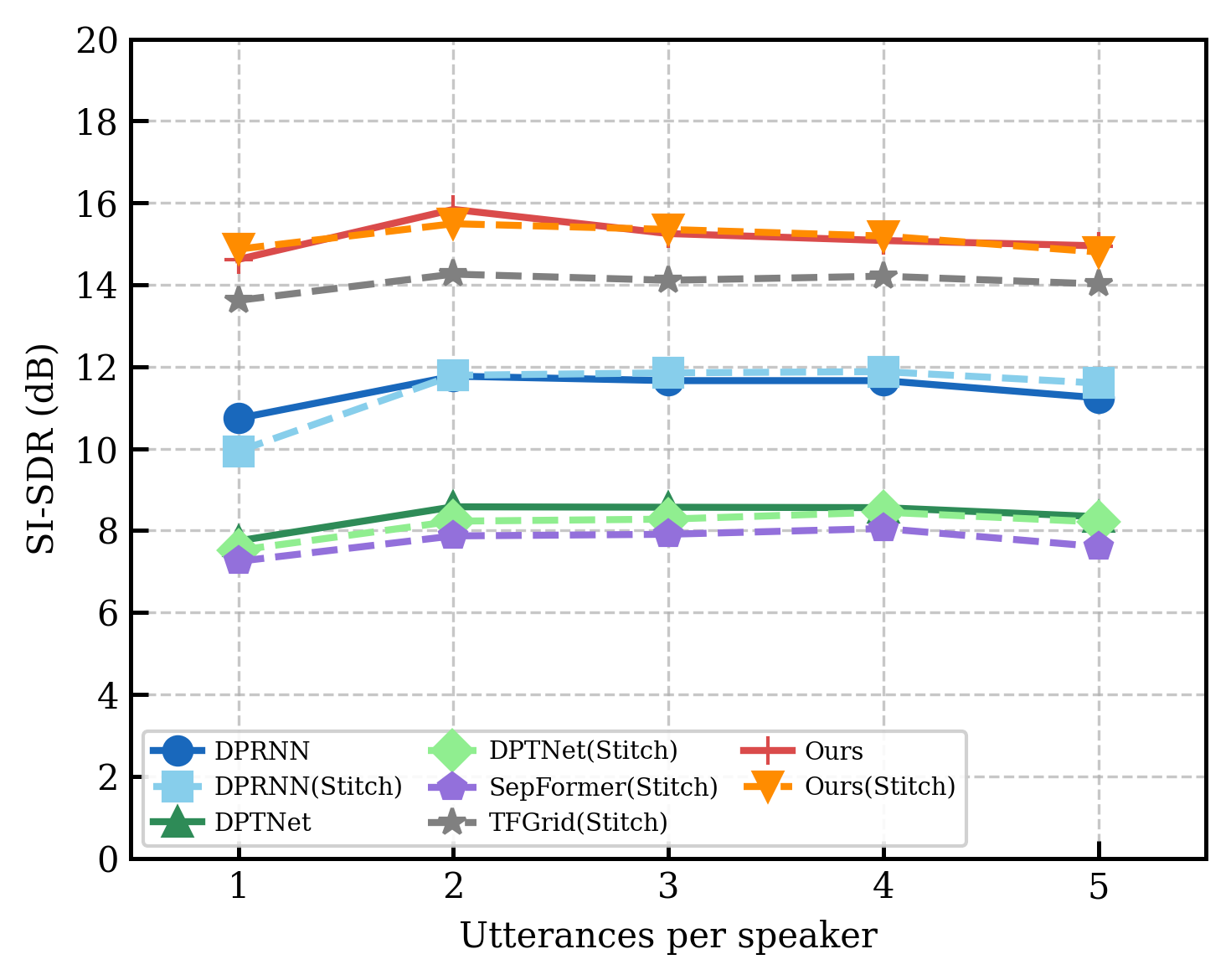}
% \vspace{-.5em}
\caption{SI-SDR performance comparison across different utterances per speaker (test sets \# $1$ and \# $6$-$9$ in Table~\ref{table_testset}). The average utterance gap is $3$~s.}
\label{fig-utt-num}
% \vspace{-.5cm}
% \vspace{-.4cm}
\vspace{2ex}
\end{figure}

As shown in Fig.~\ref{fig-utt-num}, our proposed model consistently outperforms all baselines for direct inference across different utterance counts. It achieves $14.6$~dB SI-SDR for single-utterance scenarios and reaches its peak performance of $15.8$~dB with two utterances per speaker. Even as the utterance count increases to five, our model maintains robust performance at $15.0$~dB, exhibiting only a $0.8$~dB decrease from its peak.
DPRNN demonstrates relatively stable performance, reaching $11.8$~dB with two utterances and maintaining $11.2$~dB with five utterances ($0.6$~dB decrease). DPTNet shows consistent performance across different utterance numbers, with variations less than $0.2$~dB, maintaining around $8.5$~dB.

With oracle stitching, DPRNN improves to $9.9$-$11.9$~dB range, DPTNet to $7.5$-$8.5$~dB range, and TFGrid achieves $13.6$-$14.3$~dB. SepFormer maintains $7.3$-$8.1$~dB. Our model achieves $14.5$-$15.4$~dB.
Notably, the performance gap between direct inference and oracle stitching across all models is relatively small across all utterance counts. 
This reveals that increasing the number of utterances (and consequently signal length) does not significantly increase separation difficulty or permutation error when utterance gaps are moderate. 
Such observation contrasts with the performance degradation seen in longer gap scenarios (Fig.~\ref{fig-utt-gap}), suggesting that utterance gap duration has a greater impact on separation performance than the number of utterances.

The above comparison demonstrates that FTRNN not only shows a significant performance advantage, but also achieves direct inference performance approaching or even exceeding the performance limit under ideal permutation.

% \vspace{-1ex}
% \vspace{-1mm}
\vspace{2ex}
\section{Conclusions}
\label{sec-conclusions}
\vspace{2ex}
This paper investigated how speech separation methods perform in multi-utterance recordings when trained on short segments. 
Our analysis revealed that direct inference on long signals outperformed segment-separation-stitch results even with ideal permutation information, as operating on longer signals enables the models to capture extended temporal context. 
Furthermore, experimental results revealed that utterance gap duration has a greater impact on separation performance than the number of utterances.
The proposed FTRNN trained on $10$-second segments can process unseen longer inputs ($21$-$121$~s) without segmentation while maintaining speaker association across utterance gaps that exceed training conditions, eliminating boundary distortions in conventional segment-separation-stitch approaches. 
Future work could explore extending this approach to scenarios with more than two speakers and multi-channel settings.

\newpage
\bibliographystyle{IEEEtran}
\bibliography{Bib_YWang}

\end{document}